\documentclass[a4paper]{jpconf}
\usepackage{graphicx}
\def\Jninth{PSR~J0034$-$0534 }

\def\Jfirst{PSR~J1939+2134 }

\def\JfirstC{PSR~J1939+2134, }
\def\JfirstF{PSR~J1939+2134. }
\def\Jblack{PSR~J1959+2048 }

\def\JRansom{PSR~J2214+3000 }
\def\gaeq{\raise.2ex\hbox{$>$}\kern-.75em\lower.9ex\hbox{$\sim$}\,}
\def\laeq{\raise.2ex\hbox{$<$}\kern-.75em\lower.9ex\hbox{$\sim$}\,}

\begin{document}
\title{What can we learn from phase alignment of $\gamma$-ray and radio pulsar light curves?}

\author{C Venter$^1$, T~J Johnson$^{2,3}$ and A~K Harding$^2$}
\address{$^1$~Centre for Space Research, North-West University, Potchefstroom Campus, Private Bag X6001, Potchefstroom 2520, South Africa}
\address{$^2$~Astrophysics Science Division, NASA Goddard Space Flight Center, Greenbelt, MD 20771, USA}
\address{$^3$~Department of Physics, University of Maryland, College Park, MD 20742, USA}
\ead{Christo.Venter@nwu.ac.za}

\begin{abstract}
The \textit{Fermi} Large Area Telescope (LAT) has revolutionized high-energy (HE) astronomy, and is making enormous contributions particularly to $\gamma$-ray pulsar science. As a result of the many new pulsar discoveries, the $\gamma$-ray pulsar population is now approaching~100. Some very famous millisecond pulsars (MSPs) have also been detected: J1939+2134 (B1937+21), the first MSP ever discovered, as well as J1959+2048 (B1957+20), the first black widow pulsar system. These, along with other MSPs such as PSR~J0034$-$0534 and J2214+3000, are rare among the pulsar population in that they exhibit nearly phase-aligned radio and $\gamma$-ray light curves (LCs). Traditionally, pulsar LCs have been modelled using standard HE models in conjunction with low-altitude conal beam radio models. However, a different approach is needed to account for phase-aligned LCs. We explored two scenarios: one where both the radio and $\gamma$-ray emission originate in the outer magnetosphere, and one where the emission comes from near the polar caps (PCs) on the stellar surface. We find best-fit LCs using a Markov chain Monte Carlo (MCMC) technique for the first class of models. The first scenario seems to be somewhat preferred, as is also hinted at by the radio polarization data. This implies that the phase-aligned LCs are possibly of caustic origin produced in the outer magnetosphere, in contrast to the usual lower-altitude conal beam radio models. Lastly, we constrain the emission altitudes with typical uncertainties of $\sim10\%$ of the light cylinder radius. The modelled pulsars are members of a third $\gamma$-ray MSP subclass, in addition to two others with non-aligned radio and $\gamma$-ray LCs. 
\end{abstract}

\section{Introduction}
For quite some time, there were only a handful of pulsars detected in the $\gamma$-ray waveband~\cite{Thompson04}. This small sample already provided a starting point for spectral and light curve (LC) modelling as well as preliminary population studies. However, the launch of the \textit{Fermi} Large Area Telescope (LAT)~\cite{Atwood09} heralded an exciting new era, particularly for high-energy (HE) pulsar physics. 



\subsection{$\gamma$-ray pulsars galore}
Following the first \textit{Fermi} Catalog which included about 1~500 $\gamma$-ray sources~\cite{Ferim_1st_Cat}, the second \textit{Fermi} Catalog is in now production, containing nearly 1~900 $\gamma$-ray sources~\cite{Fermi_2nd_CatA,Fermi_2nd_CatB} including almost~100 $\gamma$-ray pulsars~\cite{PSR_cat10,Smith11}. About a third of these are millisecond pulsars (MSPs), some having been found by observing non-variable, unassociated \textit{Fermi} sources at high latitudes with radio telescopes and searching for pulsed radio signals~\cite{Cognard11,Keith11,Ransom11}. Some of these MSPs exhibit the unusual phenomenon that the peaks of their $\gamma$-ray and radio LCs occur at the same observer phase (`longitude' $\phi$), i.e., they are phase-aligned.

\subsection{Traditional models}
Two classes of models have been invoked to explain HE pulsar radiation. The first class, low-altitude polar cap (PC) models~\cite{Daugherty82,DH96}, assume that primary electrons are accelerated above the neutron star surface, and that magnetic pair production of curvature radiation or inverse-Compton-scattered $\gamma$-rays occurs in the intense B-fields close to the stellar surface. In the case that pair creation is suppressed along the last open field lines, which mark the boundary between the corotating closed field line zone and the active open one, a slot gap (SG)~\cite{Arons83} may  form, corresponding to a two-pole caustic (TPC) geometry~\cite{Dyks03} which extends from the stellar surface up to near the light cylinder. The second class of models are the outer gap (OG) models~\cite{CHR86a,Romani96}, which assume the production of HE radiation along the last open field lines above the so-called null charge surface, where the Goldreich-Julian charge density becomes zero. In both classes of outer-magnetospheric models (SG and OG), the HE pulse profiles are the result of the formation of \textit{caustics} -- the accumulation of photons in narrow phase bands due to a combination of special relativistic effects and magnetic field line curvature. Also, the narrow gaps in these models require abundant production of electron-positron pairs which will be able to screen the accelerating E-field. 

For the case of MSPs, LCs and spectra have been modelled~\cite{Frackowiak05,HUM05,VdeJ05} using a pair-starved polar cap (PSPC) model \cite{MH04_PS}. The latter is similar to the traditional PC model, but in this case the number of pairs is not sufficient to screen the accelerating E-field, such that the accelerating region above the star is pair-starved. MSP spectra and energetics have also been modelled using an OG model~\cite{Zhang03,Zhang07}, while LC modelling using an annular gap model can furthermore reproduce the salient features of the $\gamma$-ray LCs~\cite{Du10}.

The outer magnetospheric models usually invoke relatively low-altitude `core' and `conal' radio beams centred on the magnetic axis, in addition to the HE radiation being produced in extended regions reaching up to the light cylinder. The difference in the location of the $\gamma$-ray and radio emission regions thus implies that the corresponding pulse profiles will have non-zero lags (phase offsets) between the $\gamma$-ray and radio LCs.

\subsection{Phase-aligned radio and $\gamma$-ray light curves: a new MSP subclass}
The LCs of the first~8 \textit{Fermi}-detected $\gamma$-ray MSPs~\cite{Abdo09_MSP} have been modelled~\cite{Venter_MSP09}, yielding two distinct MSP subclasses: those whose LCs are well fit by a standard OG or TPC model, and those whose LCs are well fit by a PSPC model. These fits are mutually exclusive. Importantly, this unexpectedly implied that there is copious pair production taking place even in MSP magnetospheres with their characteristic low B-fields. 

A third class of MSPs emerged with the discovery of PSR~J0034$-$0534~\cite{Abdo10_J0034}, which exhibited (nearly) phase-aligned radio and $\gamma$-ray LCs. Prior to this detection, such behaviour has only been observed for the Crab pulsar~\cite{Kniffen74}. This phenomenon has now also been seen for \JfirstC \Jblack~\cite{Guillemot11_2msps}, and \JRansom~\cite{Ransom11}. Some other MSPs, including PSR~B1821$-$24 and PSR~J0737$-$3039A, may also have phase-aligned radio and $\gamma$-ray LCs, on the basis that their radio and X-ray profiles are phase-aligned~\cite{Backer97,Chatterjee07}. 

\section{Geometric models predicting phase-aligned light curves}
The non-zero radio-to-$\gamma$ lags of MSPs with phase-aligned LCs disqualify the application of standard models, since the phase alignment implies co-located emission regions. We consider two model scenarios, one where the $\gamma$-rays and radio both occur in extended, high-altitude regions, and one where they originate at low altitudes. (For more details, see~\cite{Venter_Aligned11}.) 

\subsection{High-altitude models}
We use the same framework as previously~\cite{Venter_MSP09}, except that the radio emission region is now extended in altitude, and not assumed to be coming from a radio cone at a single altitiude. We furthermore limit the minimum and maximum radii of the radio and $\gamma$-ray emission regions, and assume constant emissivity as a function of altitude. These models are therefore called altitude-limited OG (alOG) and TPC (alTPC) models. Technically, the traditional OG model is a specific instance of an alOG model, and similar for the traditional TPC and alTPC models.

\subsection{Low-altitude models}
The so-called low-altitude Slot Gap (laSG) models represent an alternative possibility assuming a non-caustic origin of the emission. These are actually very-low-altitude geometric SG models resembling a hollow-cone beam close to the stellar surface. We modulate the emissivity as motivated by detailed radiation models, where the emissivity rises and falls exponentially along the B-field lines, peaking at a distance of about one or two stellar radii. Additionally, we also investigated the case where the emissivity is constant.

\section{Finding optimal LC solutions in multidimensional phase space}
The models described above involve multiple free parameters describing the emission region extent and width, as well as the pulsar geometry (inclination and observer angles $\alpha$ and $\zeta$).
In order to pick statistically the best-fit parameters when comparing to the \textit{Fermi} LC data, we have developed a Markov chain Monte Carlo (MCMC) maximum likelihood procedure~\cite{Johnson11} and applied this to the alOG and alTPC models for three MSPs as indicated below (see \Tref{tab1}). The $\gamma$-ray LCs are fit using Poisson likelihood while the radio LCs are fit using a $\chi^{2}$ statistic. These two values are then combined. Using this method, we can also derive uncertainties on the minimum and maximum altitudes of the emission regions.

\section{Results}
\begin{table}[b]
\caption{\label{tab1} Best-fit pulsar geometries for different models.}
\begin{center}
\begin{tabular}{lllllll}
\br
  &\multicolumn{2}{l}{\Jninth} & \multicolumn{2}{l}{\Jfirst} & \multicolumn{2}{l}{\Jblack}\\
\br
Model  & $\alpha$ & $\zeta$ & $\alpha$ & $\zeta$ & $\alpha$ & $\zeta$\\
\mr
alOG   & 12$^\circ$ & 69$^\circ$ & 84$^\circ$ & 84$^\circ$ & 16$^\circ$ & 82$^\circ$\\
alTPC  & 30$^\circ$ & 70$^\circ$ & 75$^\circ$ & 80$^\circ$ & 43$^\circ$ & 44$^\circ$\\
laSG1  & 10$^\circ$ & 34$^\circ$ & 30$^\circ$ & 32$^\circ$ & 20$^\circ$ & 43$^\circ$\\
laSG2  & 10$^\circ$ & 37$^\circ$ & 35$^\circ$ & 25$^\circ$ & 25$^\circ$ & 45$^\circ$\\
\br
\end{tabular}
\end{center}
\end{table}

\subsection{The alOG and alTPC models}
Increasing the lower limit of the emission region's altitude while keeping the upper limit fixed has the effect of growing the PC size (i.e., deleting a ring of radiation around the PC in the phaseplot, which indicates relative intensity per solid angle vs.\ $\zeta$ and $\phi$). This suppresses low-level off-peak emission, especially for the alTPC case, so that the peaks become sharper. Conversely, decreasing the maximum radius while keeping the lower one fixed constrains the emission to a ring-like structure around the PC. As a result, the relative peak heights change in the corresponding LCs, and low-level features may become more prominent. 

\begin{figure}[t]
\begin{center}
\includegraphics[width=26pc]{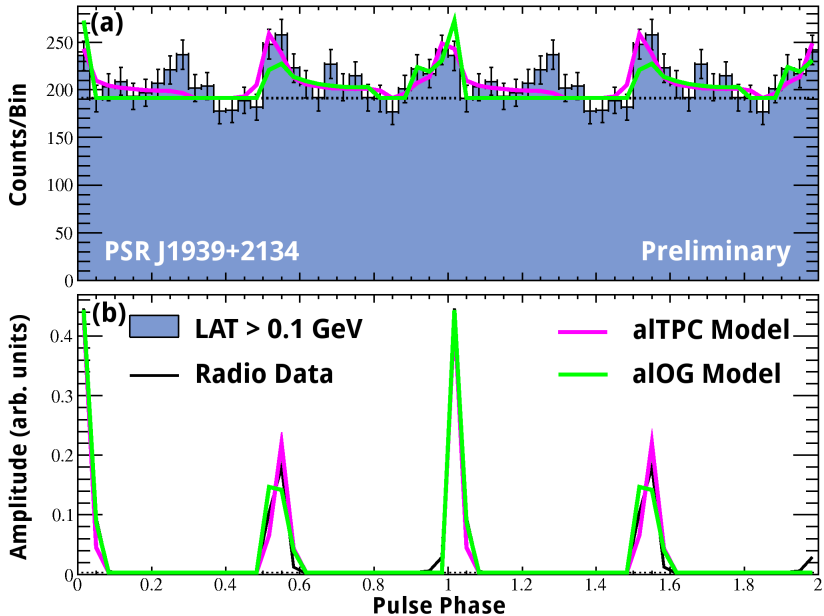}
\end{center}
\caption{\label{fig1}LC fits for \Jfirst using alTPC and alOG models. Panel~(a) shows the $\gamma$-ray data~\cite{Guillemot11_2msps}, while panel~(b) shows the radio data.}
\end{figure}

Figure~\ref{fig1} shows an example of best alOG and alTPC LC fits for \JfirstF The best-fit values for $\alpha$ and $\zeta$, for the three MSPs we have modelled, are summarized in \Tref{tab1}, having typical errors of $\sim5-10^\circ$. Best-fit emission altitudes and gap widths are presented elsewhere~\cite{Venter_Aligned11}.

\subsection{The laSG models}
It is interesting that in the laSG models the leading peak is wider while the trailing peak is sharper, even for the low-altitude emission considered here. This means that the caustic effects start to appear, given the large corotation speed of MSPs already near their surface. By including emission from higher altitudes, the peaks become broader. Also, the peak phase separation and width may be altered by choosing different values for the minimum and maximum emission altitudes. The constant-emissivity assumption leads to block-shaped LCs, and is therefore not considered viable. For the modulated-emissivity case, the peak widths may be fine-tuned using fading parameters, while the peak separation depends sensitively on $\zeta$. This class of models shows less variation in profile shape than the high-altitude ones, and there are also multiple combinations of the free parameters that give very similar LCs, so a `best solution' is probably not unique. The different best-fit solutions usually are a trade-off between best fits for $\gamma$-ray vs.\ radio LCs. For this reason, we indicate two similar laSG solutions (labeled `laSG1' and `laSG2', differing only in fading parameters which set the emissivity fading properties) in \Tref{tab1}. 

\section{Discussion}
Pulsars with radio and $\gamma$-ray LCs which are non-aligned in phase may be modelled by $\gamma$-ray emission regions extended over a large range of altitudes (OG / TPC / PSPC models) in conjunction with a conal radio beam at relatively lower emission altitudes. However, phase-aligned LCs require co-located $\gamma$-ray and radio emission regions. We studied the viability of reproducing such phase-aligned LCs using both high-altitude and low-altitude geometric models. We found that both classes of models could produce LC fits that can capture the most prominent features of the pulse profiles.

\subsection{Model constraints}
We have found best fits for the two angles describing the pulsar geometry, $\alpha$ and $\zeta$, as indicated in \Tref{tab1}. These have typical errors of $\sim5-10^\circ$. In addition, we were able to limit the emission geometry for the high-altitude models, and found that the typical errors on the best-fit emission altitudes are $\sim10\%$ of the light cylinder radius. Furthermore, the maximum radio emission altitude seems to be better constrained than the $\gamma$-ray one. Since the low-altitude models can produce very similar LCs for similar values of its free parameters, we regard the inferred values of these free parameters when applying the model to the data as less robust than the case of the altitude-limited models. 

\subsection{Which model is preferred?}
While we used the MCMC technique to search for the optimal set of parameters that produced the closest match to the LC data for the case of the altitude-limited models, we have not yet implemented this technique for the laSG models. Since the parameter space has not been fully explored for the latter class of models, we cannot really claim to have found the best fit for the alSG models. It is therefore difficult to quantitatively favour one class of models above the other. However, we did calculate the likelihood of the manually-selected best-fit laSG LCs as well as the MCMC-selected best-fit LCs of the altitude-limited models, and found that the altitude-limited models give better fits than the laSG ones, while the alTPC models provide slightly better LC fits than the alOG models. This implies that the phase-aligned $\gamma$-ray and radio LCs are most probably of caustic origin, produced in the outer magnetosphere, and the radio emission is most likely originating near the light cylinder (see also~\cite{Manchester05}). 

\subsection{Does causticity imply phase-alignment?}
By studying the visibility of two samples of $\gamma$-ray and radio pulsars, it was concluded~\cite{Ravi10} that radio and $\gamma$-ray beams must have comparable sky coverage, especially for pulsars with large spin-down luminosities. This implies that the radio emission should originate in very wide beams at a significant fraction of the light cylinder, motivating studies of high-altitude caustic radio emission. However, this scenario is limited in application to pulsars having nearly phase-aligned LCs (with phase differences of up to $\sim0.2$), unless the radio emission region is significantly offset from the $\gamma$-ray one. The bulk of the $\gamma$-ray pulsar population exhibits quite large ($\sim0.1-0.5$) radio-to-$\gamma$-ray phase lags, so that caustic radio emission is probably is not ubiquitous in young pulsars. Furthermore, the $\gamma$-ray profiles are usually double-peaked, while the radio ones are single-peaked. Unless the radio emissivity has a strong altitudinal or azimuthal dependence, both radio and $\gamma$-ray LCs would be mostly double-peaked if they result from caustic emission, which is not observed. On the other hand, radio caustics may be more common for short-period MSPs, as there are many more examples of MSPs with phase-aligned LCs. Detailed population studies involving both young and old pulsars will provide more quantitative answers to this question.

\subsection{The potential of polarization}
The rotating vector model~\cite{RC69} generally provides a good description of radio polarization data (position angle as function of phase). However, this model assumes a static dipole B-field with no rotation effects included, and is probably not valid for fast-spinning MSPs which are expected to have significant B-field distortion. Caustic emission models predict rapid position angle swings with phase, coupled with depolarization~\cite{Dyks04_B}, since the emission from a large range of altitudes and B-field orientations is compressed into a narrow phase interval to form the peaks. Rapid changes in the polarization angle and low levels of linear and circular polarization near the peak phases have indeed been seen for the MSPs modelled in this paper, and may be indicative of caustic effects. Polarization signatures will therefore be important to help discriminate between LCs produced by caustic emission (such as occurs in alOG / alTPC models) or not (e.g., in the laSG model).

\section{Conclusions}
Future studies should include development of full radiation models which will be able to reproduce both the multiwavelength LC shapes, polarization properties, as well as the energy-dependent behaviour of the spectra of the $\gamma$-ray pulsars. Ways in which to induce abundant pair creation in low-B MSP magnetospheres, such as using offset-dipole~\cite{HM11} or multipole~\cite{Zhang03} B-fields, will need to be investigated.

\ack
CV is supported by the South African National Research Foundation. AKH acknowledges support from the NASA Astrophysics Theory Program. CV, TJJ, and AKH acknowledge support from the \textit{Fermi} Guest Investigator Program as well as fruitful discussions with Dick Manchester and Matthew Kerr.

\end{document}